\documentclass[aps, prl, twocolumn]{revtex4}
\usepackage{graphicx}
\usepackage{epsfig}
\usepackage{fancyheadings}
\usepackage{hyperref}

\newcommand{\be}{\begin{equation}}
\newcommand{\ee}{\end{equation}}

\begin{document}

\title{Torque Scaling in Turbulent Taylor-Couette Flow with Co- and Counterrotating Cylinders}

\author{Dennis P.M.\ van Gils, Sander G.\ Huisman, Gert-Wim Bruggert, Chao Sun, and Detlef Lohse}

\affiliation{Physics of Fluids Group, Faculty of Science and Technology, Impact and Mesa Institutes \& Burgers Center for Fluid Dynamics, University of Twente, 7500AE Enschede, The Netherlands}
\date{\today}

\begin{abstract}

We analyze the global transport properties of turbulent Taylor-Couette flow in the strongly turbulent regime for independently rotating outer and inner cylinders, reaching Reynolds numbers of the inner and outer cylinders of $\mathrm{Re}_i = 2 \times 10^6$ and $\mathrm{Re}_o = \pm 1.4 \times 10^6$, respectively. For all $\mathrm{Re}_i$, $\mathrm{Re}_o$,  the dimensionless torque $G$ scales as a function of the Taylor number Ta (which is proportional to the square of the difference between the angular velocities of the inner and outer cylinders) with a universal effective scaling law $G \propto \mathrm{Ta}^{0.88}$, corresponding to $\mathrm{Nu}_\omega \propto \mathrm{Ta}^{0.38}$ for the Nusselt number characterizing the angular velocity transport between the inner to the outer cylinders. The exponent $0.38$ corresponds to the ultimate regime scaling for the analogous Rayleigh-B\'enard system. The transport is most efficient for the counterrotating case along the diagonal in phase space with $\omega_o \approx -0.4 \omega_i$.
\\
\\
\noindent DOI: \href{http://dx.doi.org/10.1103/PhysRevLett.106.024502}{10.1103/PhysRevLett.106.024502}
\end{abstract}

\maketitle
\thispagestyle{fancy}

Global transport properties of turbulent flows are of prime importance for many applications of fluid dynamics, but also for a fundamental understanding, as they reflect the interplay between the boundary layer and the bulk. The two canonical systems used to analyze the transport properties in closed turbulent systems are Rayleigh-B\'enard (RB) convection and Taylor-Couette (TC) flow, and they are conceptually closely related \cite{bra69,dub02,eck07b}. In RB flow, heat (in dimensionless form the Nusselt number) is transported from the hot bottom plate to the cold top plate \cite{ahl09,loh10}, whereas in TC flow angular velocity is transported from the inner to the outer cylinder (for $\omega_i>\omega_o$). In analogy to RB flow, Eckhardt {\it et al.} \cite{eck07b} identified, from the underlying Navier-Stokes equations,
\be
J^\omega = r^3 ( \left< u_r \omega \right>_{A,t} - \nu \partial_r \left< \omega \right>_{A,t}),
\label{j_omega}
\ee
as a relevant conserved transport quantity, representing the flux of angular velocity from the inner to the outer cylinder. Here $u_r$($u_\phi$) is the radial (azimuthal) velocity, $\omega = u_\phi/r$ the angular velocity, and $\left< \dots\right>_{A,t}$ characterizes averaging over time and an area with constant $r$ from the axis. $J^\omega$ is made dimensionless with its value $J^\omega_{lam} = 2\nu r_i^2 r_o^2(\omega_i - \omega_o)/ (r_o^2 - r_i^2)$ for the laminar case, giving a ``Nusselt number'' as a dimensionless transport quantity,
\be
\mathrm{Nu}_\omega = J^\omega / J^\omega_{lam},
\ee
where $r_{i,o}$ and $\omega_{i,o}$ denote the radius and the angular velocity of the inner and outer cylinders, respectively, and $\nu$ is the kinematic viscosity of the fluid. Nu$_\omega$ is closely connected to the torque $\tau$ that is necessary to keep the inner cylinder rotating at a constant angular velocity or, in dimensionless form, to
\be
G = { \tau\over 2\pi \ell \rho_{\mathrm{fluid}} \nu^2 } = \mathrm{Nu}_\omega {J^\omega_{lam}\over  \nu^2 } = \mathrm{Nu}_\omega G_{lam},
\label{g}
\ee
where $\ell$ is the height of the cylinder and $\rho_{\mathrm{fluid}}$ the density of the fluid. Yet another often used possibility to represent the data is the friction coefficient $c_f = [(1-\eta)^2/\pi]G/\mathrm{Re}_i^2$ \cite{lat92a}.

\begin{figure*}[ht]
\begin{center}
\epsfig{file=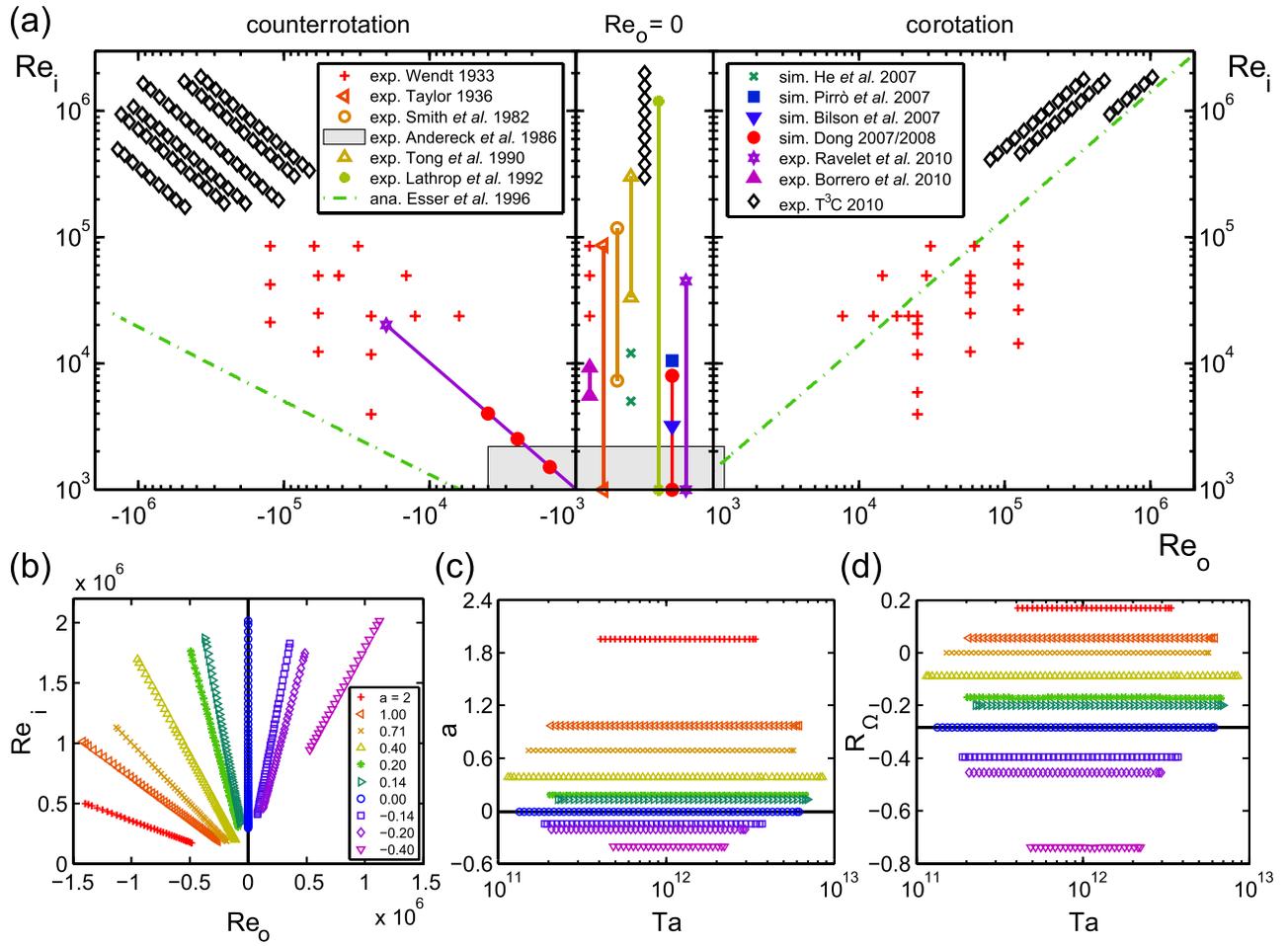, width = 17cm}
\caption{(a) Explored phase space (Re$_o$, Re$_i$) of TC flow with independently rotating inner and outer cylinders. To the right of the horizontal axis the cylinders are corotating, to the left of it they are counterrotating, and a log-log representation has been chosen. Experimental data by
Wendt \cite{wen33} (pluses),
Taylor \cite{tay36} (left triangles),
Smith and Townsend \cite{smi82} (open circles),
Andereck {\it et al.} \cite{and86} (grey box),
Tong {\it et al.} \cite{ton90} (upward triangles),
Lathrop {\it et al.} \cite{lat92a} (stars),
Ravelet {\it et al.} \cite{rav10} (hexagrams),
Borrero-Echeverry {\it et al.} \cite{bor10} (upward solid triangles),
and simulations by
Pirro and Quadrio \cite{pir08} (solid squares),
Bilson and Bremhorst \cite{bil07} (downward solid triangles),
and
Dong \cite{don07, don08} (solid circles).
The dashed lines are Esser and Grossmann's \cite{ess96} estimate for the onset of turbulence with $\eta = 0.71$. The many data points in the small Reynolds number
regime of pattern formation and spatial temporal chaos (see e.g.\ \cite{and86,pfi81,cro93}) have not been included in this phase diagram.
Our data points for this publication are the black diamonds. Our data points for this publication are the black diamonds. (b) Our data points in the phase diagram on a linear scale. (c) Our data points in the phase space (Ta, $a$); note that Ta also depends on $a$. (d) Our data points in the phase space (Ta, $R_{\Omega}$) [see Eq.\ (\ref{rotnum})].
}
\label{fig1}
\end{center}
\end{figure*}

For RB flow, the scaling properties of the Nusselt number in the fully turbulent regime (i.e., for very large Rayleigh numbers, say Ra $\ge 10^{10}$) have received tremendous attention in the last decade and various heat flux measurements have been performed; again, see the review article \cite{ahl09}. In contrast, TC flow in the fully turbulent regime has received much less attention, with the only exception being the Texas experiment by Swinney, Lathrop, and coworkers \cite{lat92,lat92a,lew99,ber03}. In that experiment a Reynolds number Re$_i = 10^6$ of the inner cylinder was reached (with the outer cylinder at rest) and an effective power law of $G\propto \mathrm{Re}_i^{\gamma}$ with $\gamma \approx 1.6 - 1.86$ was detected \cite{lat92,lat92a} in the turbulent regime, though the scaling properties are not particularly good and, strictly speaking, $\gamma$ depends on Re$_i$; i.e., there is no pure scaling. Indeed, in Refs.\ \cite{eck00,eck07a,eck07b} we have argued that there should be a smooth transition from $\gamma = 3/2$ for the small Reynolds number of a boundary layer dominated flow to $\gamma = 2$ for the larger Reynolds number of a flow dominated by the turbulent bulk. Turbulent TC experiments get even more scarce for TC flow with inner and outer cylinders rotating independently. We are only aware of the Wendt experiments in the 1930s \cite{wen33}, reaching Re$_i \approx 10^5$ and Re$_o \approx \pm 10^5$ and the recent ones by Ravelet {\it et al.} \cite{rav10}, reaching Re$_i \approx 5 \times 10^4$ and Re$_o = \pm 2 \times 10^4$.
The hitherto explored phase diagram of TC flow with independently rotating cylinders is shown in Fig.\ \ref{fig1}(a). An alternative representation of the phase diagram is given by Dubrulle {\it et al.} \cite{dub05}, who introduce a shear Reynolds number and a rotation number as an alternative representation of the phase space (see below).

\begin{figure}[hb]
\begin{center}
\epsfig{file=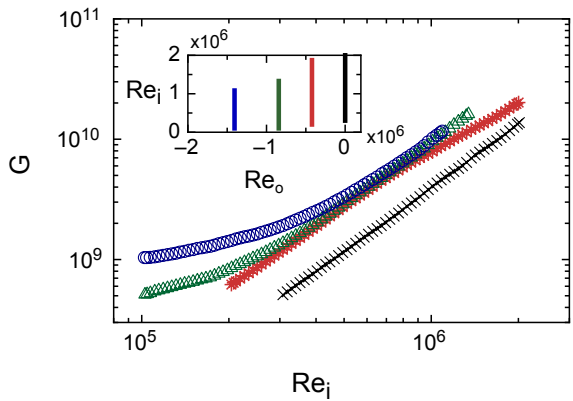, width = 7.5cm}
\caption{The dimensionless torque $G(\mathrm{Re}_i)$ for counterrotating TC flow for four different fixed values of Re$_o =  -1.4 \times 10^6$, $-0.8 \times 10^6$, $-0.4 \times 10^6$, and $0$ (top to bottom data sets); see inset for the probed area of the parameter space.
}
\label{fig2}
\end{center}
\end{figure}

In the phase diagram [Fig.\ \ref{fig1}(a)] we have also added the Re$_i$, Re$_o$ numbers that we explored with our newly constructed Twente turbulent TC facility ($T^3C$), which we have described in great detail in Ref.\ \cite{gil11a}. In short, at this facility the cylinders are $\ell = 0.97$ m high; the inner cylinder has a radius of $r_i = 20$ cm, the outer cylinder has a radius of $r_o = 27.9$ cm, and the maximal inner and outer angular velocity are $\omega_i/2\pi = 20$ Hz and $\omega_o/2\pi = \pm 10$ Hz, respectively, corresponding to Re$_i = r_i \omega_i d / \nu = 2\times 10^6$ and Re$_o = r_o \omega_o d/ \nu = \pm 1.4 \times 10^6$, with $d = r_o - r_i$. The system is fully temperature controlled through cooling of the upper and lower plates. The torque is  measured at the middle part of the inner cylinder (similar to \cite{lat92}) by load cells imbedded inside the inner cylinder and not by measuring the torque through the seals. One of the goals we want to achieve with this new facility is to explore the (Re$_i$, Re$_o$, $\eta = r_i/r_o$) parameter space, thus entering {\it terra incognita}, and measure the torque [i.e., transport of the angular velocity (Nu$_\omega$ in dimensionless form) or, again expressed differently, the overall drag] and the internal Reynolds number of the flow.

In this Letter we will focus on the required torque for fully developed turbulent flow (Re$_i$, Re$_o > 10^5$), where $\eta = 0.716$ with independently rotating inner and outer cylinders, which hitherto has not been explored. The examined parameter space in this Letter is shown in the space of (Re$_i$, Re$_o$) in Fig.\ 1(b), (Ta, $a$) in Fig.\ 1(c), and (Ta, $R_{\Omega}$) in Fig.\ 1(d), to be explained below. We will not address the question whether pure scaling laws exist: First, the explored Reynolds number range is too short to answer this question, and second, the earlier work \cite{lat92,lat92a,lew99,eck00,eck07a,eck07b} gives overwhelming experimental and theoretical evidence that there are no pure scaling laws, even up to Reynolds numbers of $10^6$. So all scaling exponents in this Letter have to be read as effective scaling laws.

Our results for the counterrotating case for the dimensionless torque $G$ as function of Re$_i$ for fixed Re$_o$ are shown in Fig.\ \ref{fig2}. One immediately sees that counterrotation enhances the torque (and thus the overall drag), but that for general Re$_o\ne 0$ the effective power law $G\propto \mathrm{Re}_i^{1.76}$, that holds in the case of inner cylinder rotating only, gets lost; in fact, there is even no effective power law at all.

How can we represent the data to better reveal the transport properties of the system? The analysis of Eckhardt {\it et al.} \cite{eck07b} and the analogy of the TC system to the RB system suggest to better plot Nu$_\omega$ as a function of the Taylor number
\be
\mathrm{Ta} = \frac{1}{4} \sigma {d^2 (r_i + r_o)^2 (\omega_i - \omega_o)^2 \nu^{-2}},
\label{ta}
\ee
where $\sigma = \{[(1 + \eta)/2]/\sqrt{\eta}\}^4$, i.e., along the diagonals
\be
\omega_o = -a \omega_i
\label{dia}
\ee
in the parameter space \cite{note1}, Fig.\ \ref{fig1}(b). Indeed, Eckhardt {\it et al.}\ \cite{eck07b} derived, from the underlying Navier-Stokes equation, the exact relation
\be
\epsilon_w = \epsilon - \epsilon_{lam} = \nu^3 d^{-4} \sigma^{-2} \mathrm{Ta} (\mathrm{Nu}_\omega - 1)
\label{epsw}
\ee
for the excess kinetic energy dissipation rate $\epsilon_{w}$ [i.e., the total kinetic energy dissipation rate $\epsilon$ minus the kinetic energy dissipation rate in the laminar case $\epsilon_{lam} = 4\nu r_i^2 r_o^2 (r_i + r_o)^{-2} d^{-2} (\omega_i - \omega_o)^2$]. In Eq.\ (\ref{epsw}) $\sigma$ can be interpreted as a (geometric) Prandtl number, and Ta and Nu$_\omega$ are the exact TC analogs to the Rayleigh and Nusselt numbers in RB flow. Along the diagonal, Eq. (\ref{dia}), in parameter space, one has Ta $= {1\over 4} \sigma d^2 (r_i + r_o)^2 (1 + a)^2 \omega_i^2 \nu^{-2}$, and the well-studied \cite{ahl09} effective scaling law Nu $\propto$ Ra$^{\tilde \gamma}$ for RB flow (with $\tilde \gamma \approx 0.31$ \cite{nie00,ahl09}) would now correspond to an effective scaling law Nu$_\omega \propto \mathrm{Ta}^{\gamma}$ for TC flow.

\begin{figure}[hb]
\begin{center}
\epsfig{file=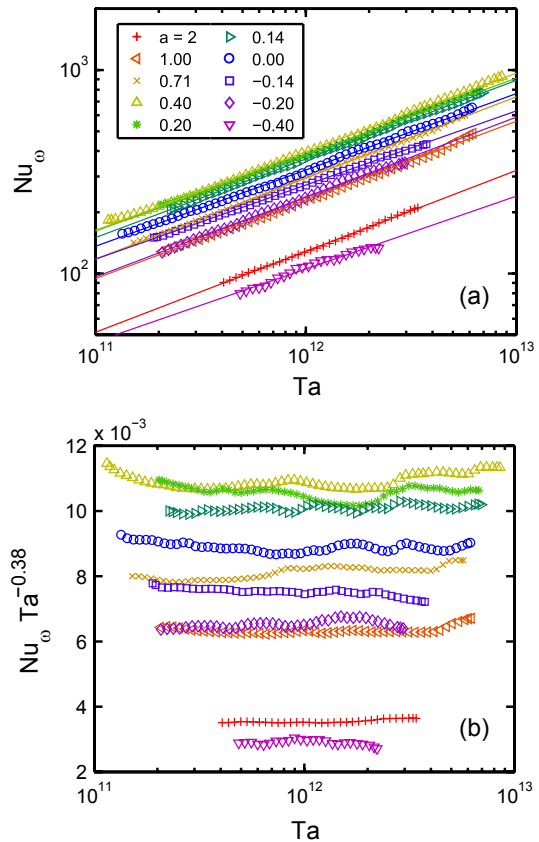, width = 7cm}
\caption{(a) Nu$_\omega$ vs Ta for various $a$; see Fig.\ \ref{fig1}(b) for the location of the data in parameter space. A universal effective scaling Nu$_\omega \propto \mathrm{Ta}^{0.38}$ is revealed. The compensated plots Nu$_\omega / \mathrm{Ta}^{0.38}$ in (b) show the quality of the effective
scaling and the $a$-dependent prefactor of the scaling law.
}
\label{fig3}
\end{center}
\end{figure}

Nu$_\omega$ vs Ta is shown in Fig.\ \ref{fig3}(a) for various $a$, i.e., along
various straight lines through the origin of the parameter space, Fig.\ \ref{fig1}(b). A {\it universal}, i.e.\ $a$-indepedent, effective scaling Nu$_\omega \propto \mathrm{Ta}^{\gamma}$ with $\gamma \approx 0.38$ is clearly revealed. This corresponds to a scaling of $G\propto \mathrm{Re}_i^{1.76}$ for the dimensionless torque along the straight lines, Eq. (\ref{dia}), in the parameter space, Fig.\ \ref{fig1}, to $c_f \propto \mathrm{Re}_i^{-0.24}$ for the drag coefficient, and to $G \propto \mathrm{Ta}^{0.88}$. The compensated plots Nu$_\omega / \mathrm{Ta}^{0.38}$ in Fig.\ \ref{fig3}(b) demonstrate the quality of the effective scaling and, in addition, show the $a$ dependence of the prefactor of the scaling law.

\begin{figure}[htbp]
\begin{center}
\epsfig{file=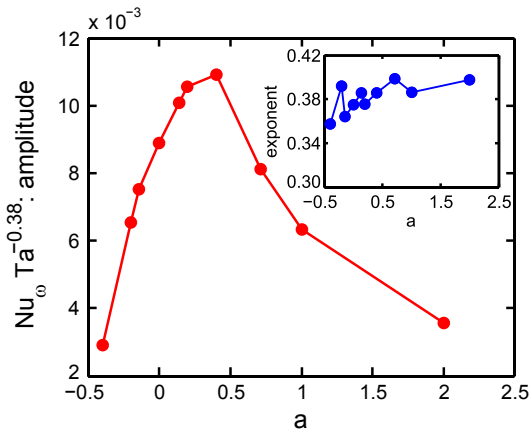, width = 7cm}
\caption{Prefactor of the effective scaling law Nu$_\omega \propto \mathrm{Ta}^{0.38}$ (shown in Fig.\ \ref{fig3}) as a function of $a = -\omega_o/\omega_i$. The inset shows the effective exponents $\gamma$ which results from an individual fit of the scaling law Nu$_\omega \propto \mathrm{Ta}^{\gamma}$.
}
\label{fig4}
\end{center}
\end{figure}

The $a$ dependence of the prefactor Nu$_\omega /\mathrm{Ta}^{0.38}$ is plotted in Fig.\ \ref{fig4}. It shows a pronounced maximum around $a = 0.4$, i.e., for the
moderately counterrotating case, signaling the most efficient angular velocity transport from the inner to the outer cylinder at that value. We mention that it is obvious that this curve has a maximum, as in both limiting cases $a\to \pm \infty$ (rotating of the outer cylinder only) the flow is laminar and Nu$_\omega = 1$, but it is interesting to note that the maximum does not occur for the most pronounced counterrotating case $\omega_o = -\omega_i$ (or $a = 1$). Compared to the case of pure inner cylinder rotation ($a = 0$), at $a = 0.4$ the angular
velocity transport from the inner to the outer cylinder is enhanced by more than 20\%.

The parameter $a = -\omega_o/\omega_i$ is connected to the so-called rotation number
\be
R_\Omega = (1 - \eta)(\mathrm{Re}_i + \mathrm{Re}_o)/(\eta \mathrm{Re}_o - \mathrm{Re}_i)
\label{rotnum}
\ee
introduced by Dubrulle {\it et al.} \cite{dub05} and used by Ravelet {\it et al.} \cite{rav10}, namely, $R_\Omega = (1 - \eta)\eta^{-1} (a - \eta)(a + 1)^{-1}$. We also plot our data points in the phase space of (Ta, $R_{\Omega}$), as shown in Fig.\ 1(d). The optimal value $a\approx 0.4$ we found for the transport properties of the system corresponds to $R_\Omega \approx  -0.09$. In this Letter we prefer $a$ as compared to $R_\Omega$, as the sign of $a$ immediately signals whether the system is corotating or counterrotating.

In conclusion, we have explored the {\it terra incognita} of fully developed turbulent TC flow with independently rotating inner and outer cylinders, beyond Reynolds numbers of $10^6$, finding a universal scaling law $G\propto \mathrm{Ta}^{0.88}$, corresponding to Nu$_\omega \propto \mathrm{Ta}^{0.38}$, for all (fixed) $a = - \omega_o/\omega_i$, with optimal transport quantities at $a \approx 0.4$. It is remarkable that the effective scaling exponent 0.38 exactly resembles the analogous effective scaling exponent in Nu $\propto \mathrm{Ra}^{0.38}$ in RB convection in the ultimate regime of thermal convection \cite{cha97, cha01}, reflecting the analogy between TC and RB flow also in the strongly turbulent regime.

The next steps will be to further extend the parameter space, Fig.\ \ref{fig1}, towards further radius ratios $\eta$ to see whether the observed universality carries on towards an even larger parameter range, and to also measure the Taylor-Reynolds number and the wind Reynolds numbers of the internal flow, which are closely connected to Nu$_\omega$ and for which theoretical predictions exist \cite{eck07b}. With such measurements and characterizations of the flow structures, we will also be able to check whether these are reflected in the overall transport properties.

We thank G.\ Ahlers, B.\ Eckhardt, D.\ P.\ Lathrop, S.\ Grossmann, G.\ Pfister, and E.\ van Rietbergen for various stimulating discussions about TC flow over the years, D. Narezo Guzman for help with some measurements, TCO-TNW Twente for the collaboration on building up the facility, and STW, which is financially supported by NWO, for financial support.

\textit{Note added in proof.}---After submission of our Letter, Dan Lathrop mase us aware of the parallel work \cite{pao10}, independently confirming the peak in the dimensionless torque as a function of $a$.

\bibliographystyle{prsty}

\begin{thebibliography}{10}

\bibitem{bra69}
P. Bradshaw, J. Fluid Mech. {\bf 36}, 177 (1969).

\bibitem{dub02}
B. Dubrulle and F. Hersant, Eur. Phys. J. B {\bf 26}, 379 (2002).

\bibitem{eck07b}
B. Eckhardt, S. Grossmann, and D. Lohse, J. Fluid Mech. {\bf 581}, 221 (2007).

\bibitem{ahl09}
G. Ahlers, S. Grossmann, and D. Lohse, Rev. Mod. Phys. {\bf 81}, 503 (2009).

\bibitem{loh10}
D. Lohse and K.~Q. Xia, Annu. Rev. Fluid Mech. {\bf 42}, 335 (2010).

\bibitem{lat92a}
D. P. Lathrop, J. Fineberg, and H. L. Swinney, Phys. Rev. A {\bf 46}, 6390 (1992).

\bibitem{lat92}
D. P. Lathrop, J. Fineberg, and H. L. Swinney, Phys. Rev. Lett. {\bf 68}, 1515 (1992).

\bibitem{lew99}
G. S. Lewis and H. L. Swinney, Phys. Rev. E {\bf 59}, 5457 (1999).

\bibitem{ber03}
T. H. van den Berg, C.R. Doering, D. Lohse, and D. P. Lathrop, Phys. Rev. E {\bf 68}, 036307 (2003).

\bibitem{eck00}
B. Eckhardt, S. Grossmann, and D. Lohse, Eur. Phys. J. B {\bf 18}, 541 (2000).

\bibitem{eck07a}
B. Eckhardt, S. Grossmann, and D. Lohse, Europhys. Lett. {\bf 78}, 24001 (2007).

\bibitem{wen33}
F. Wendt, Ingenieurs-Archiv {\bf 4}, 577 (1933).

\bibitem{ton90}
P. Tong, W. I. Goldburg, J. S. Huang, and T. A. Witten, Phys. Rev. Lett. {\bf 65}, 2780 (1990).

\bibitem{rav10}
F. Ravelet, R. Delfos, and J. Westerweel, Phys. Fluids {\bf 22}, 055103 (2010).

\bibitem{dub05}
B. Dubrulle {\emph et al.}, Phys. Fluids {\bf 17}, 095103 (2005).

\bibitem{gil11a}
D. van Gils {\emph et al.}, Rev. Sci. Instrum. (to be published).

\bibitem{note1}
We define Eq.\ (5) with a minus sign, as our focus is on the counterrotating case.

\bibitem{nie00}
J. Niemela, L. Skrbek, K. R. Sreenivasan, and R. Donnelly, Nature (London) {\bf 404}, 837 (2000).

\bibitem{cha97}
X. Chavanne {\emph et al.}, Phys. Rev. Lett. {\bf 79}, 3648 (1997).

\bibitem{cha01}
X. Chavanne {\emph et al.}, Phys. Fluids {\bf 13}, 1300 (2001).

\bibitem{tay36}
G. I. Taylor, Proc. R. Soc. A {\bf 157}, 546 (1936).

\bibitem{smi82}
G. P. Smith and A. A. Townsend, J. Fluid Mech. {\bf 123}, 187 (1982).

\bibitem{and86}
C. D. Andereck, S. S. Liu, and H. L. Swinney, J. Fluid Mech. {\bf 164}, 155 (1986).

\bibitem{bor10}
D. Borrero-Echeverry, M. F. Schatz, and R. Tagg, Phys. Rev. E {\bf 81}, 025301 (2010).

\bibitem{pir08}
D. Pirro and M. Quadrio, Eur. J. Mech. B, Fluids {\bf 27}, 552 (2008).

\bibitem{bil07}
M. Bilson and K. Bremhorst, J. Fluid Mech. {\bf 579}, 227 (2007).

\bibitem{don07}
S. Dong, J. Fluid Mech. {\bf 587}, 373 (2007).

\bibitem{don08}
S. Dong, J. Fluid Mech. {\bf 615}, 371 (2008).

\bibitem{ess96}
A. Esser and S. Grossmann, Phys. Fluids {\bf 8}, 1814 (1996).

\bibitem{pfi81}
G. Pfister and I. Rehberg, Phys. Lett. {\bf 83}, 19 (1981).

\bibitem{cro93}
M.~C. Cross and P.~C. Hohenberg, Rev. Mod. Phys. {\bf 65}, 851 (1993).

\bibitem{pao10}
M. S. Paoletti and D. P. Lathrop, preceding Letter, Phys. Rev. Lett., {\bf 106}, 024501 (2011).

\end{thebibliography}

\end{document}